\documentclass[12pt]{article}

\usepackage{natbib}
\usepackage{hyperref}
\usepackage{graphicx}
\usepackage{amsmath, amssymb}
\usepackage{authblk} 
\usepackage{orcidlink} 
\usepackage{amsthm} 
\usepackage{booktabs}
\usepackage{tabularx}
\theoremstyle{remark}

\usepackage{arydshln}
\usepackage{natbib}
\usepackage{url}
\usepackage{amsmath, amssymb}
\usepackage{graphicx}
\usepackage{bbold}
\usepackage{mathbbol}
\usepackage{float}
\usepackage{xcolor}
\usepackage{bm}
\usepackage{setspace}
\usepackage{multirow}
\usepackage{lscape}
\usepackage{rotating}
\usepackage{adjustbox}
\usepackage{booktabs}
\usepackage{ragged2e}
\usepackage[caption=false]{subfig}
\usepackage{amsthm}
\usepackage{authblk} 
\usepackage[utf8]{inputenc}

\usepackage[normalem]{ulem} 

\definecolor{edicionInsercion}{RGB}{0, 128, 0}      
\definecolor{edicionEliminacion}{RGB}{255, 0, 0}    
\definecolor{edicionComentario}{RGB}{0, 0, 255}      

\newcommand{\norm}[1]{\left\lVert#1\right\rVert}

\newcommand{\bpm}{\begin{pmatrix}}
\newcommand{\epm}{\end{pmatrix}}

\def\*#1{\mathbf{#1}}
\def\+#1{\mathbb{#1}}
\def\^#1{\mathbb{#1}}
\DeclareSymbolFontAlphabet{\amsmathbb}{AMSb}%

\begin{document}


\title{Predicting Distributions of Physical Activity Profiles in the NHANES Database Using a Partially Linear Fréchet Single Index Model}

\author[1,$\ast$]{Marcos Matabuena}
\author[2]{Aritra Ghosal}
\author[3]{Wendy Meiring}
\author[4]{Alexander Petersen}

\affil[1]{Department of Biostatistics, Harvard University, \textit{mmatabuena@hsph.harvard.edu}}
\affil[2]{Department of Biostatistics, St. Jude Children's Research Hospital}
\affil[3]{Department of Statistics and Applied Probability, University of California, Santa Barbara}
\affil[4]{Department of Statistics, Brigham Young University}









\maketitle

\begin{abstract}
{Object-oriented data analysis is a fascinating and evolving field in modern statistical science, with the potential to make significant contributions to biomedical applications. This statistical framework facilitates the development of new methods to analyze complex data objects that capture more information than traditional clinical biomarkers. 
This paper applies the object-oriented framework to analyze physical activity levels, measured by accelerometers, as response objects in a regression model. Unlike traditional summary metrics, we utilize a recently proposed representation of physical activity data as a distributional object, providing a more nuanced and complete profile of individual energy expenditure across all ranges of monitoring intensity. A novel hybrid Fréchet regression model is proposed and applied to US population accelerometer data from National Health and Nutrition Examination Survey (NHANES) 2011-2014. The semi-parametric nature of the model allows for the inclusion of nonlinear effects for critical variables, such as age, which are biologically known to have subtle impacts on physical activity. Simultaneously, the inclusion of linear effects preserves interpretability for other variables, particularly categorical covariates such as ethnicity and sex. The results obtained are valuable from a public health perspective and could lead to new strategies for optimizing physical activity interventions in specific American subpopulations.}
\end{abstract}

\section{Introduction}
Medical science is living in a golden age with the expansion of the clinical paradigms of digital and precision medicine \cite{kosorok2019precision, JAVAID2022100379,Onnela2021}. In this new context, it is increasingly common to record patient data that is most faithfully represented using complex statistical objects such as probability distributions \cite{10.1093/jrsssc/qlad007,matabuena2020glucodensities} that contain enriched information compared to traditional clinical biomarkers in predictive terms. Distributional representations can be seen as natural digital functional biomarkers to analyze wearable data information. In a series of papers, the performance of the distributional representation was compared with that of existing summary metrics, providing strong evidence of their advantages in diabetes and physical activity domains \cite{10.1093/jrsssc/qlad007,matabuena2020glucodensities,Matabuena2022,cui4investigating}. Distributional representations are a direct functional extension of traditional compositional metrics \cite{biagi2019individual,park2025beyond,battelino2019clinical}
 and allow the creation of synthetic profiles over a continuum of intensities measured by wearable devices that provide an individualized profile of the patient's activity. Importantly, these representations overcome the critical limitations of compositional metrics to define specific cut-off points to categorize patient information that can introduce subjectivity and be highly dependent on the population under analysis.  

This work is motivated by the desire to uncover factors that are associated with the physical activity patterns of the American population, where these patterns are reprsented as distributional data objects. As energetic expenditure behaves nonlinearly with age \cite{https://doi.org/10.1111/j.1532-5415.2012.04153.x}, and other anthropometrical measures \cite{mehta2017physiological}, more advanced and flexible regression models are required to overcome the limitations of the linear model. Here, to provide a good balance between the advantages and disadvantages of linear and nonlinear models, the proposed model extends the partially linear model for scalar responses \cite{liang2010estimation} to the case of a distributional response object, yielding a partially linear Fréchet single index model. As in the scalar response case, this can be viewed as an extension of the recently proposed global Fréchet regression and Fréchet single index models. Furthermore, the survey weights from the complex survey design of the NHANES are incorporated into the model estimation to obtain reliable population-based results according to the composition of the US population \cite{lumley2010svy}.  

From a public health point of view, the proposed model is attractive because it elucidates the impact certain variables exert on the American population's average physical activity level profiles along the full range of accelerometer intensities. Moreover, these new findings can help to refine and plan specific health interventions that reduce the gap in physical inactivity in different US sub-populations. For example, one of the follow-up analyses conducted herein extracts clinical phenotypes of individuals to characterize the patients who are more or less active than predicted by the regression model.

The structure of the paper is as follows. Section \ref{sec:nhanes} introduces the NHANES data that will be analyzed, together with a background of the physical activity distributional representations. Section \ref{sec:algorithm}
introduces the model and an efficient, spline-based estimator. Section \ref{sec:results} reports the various analyses performed. Finally, Section \ref{sec:disc} discusses the results from a public health perspective, this paper's role in the broader statistical literature on regression models in metric spaces, and its opportunities in the medical field to analyze other complex statistical objects. 

\subsection{Contributions}

The methodological contributions of this paper will first be summarized, as well as the findings from analysis of the physical activity NHANES in regression models whose responses are the distributional physical activity representation.

\begin{itemize}
    \item To the authors' knowledge, the proposed model is the first partially linear Fréchet single index (PL-FSI) regression model for responses that are probability distributions, viewed as elements of the $L^2$-Wasserstein space. Moreover, for this particular situation, the use of splines are introduced for the first time in the Fréchet regression modeling framework. 
    \item  An efficient optimization strategy is proposed to address the complex survey sampling mechanism of the NHANES data that retains the estimator's form of a weighted least squares problem. The key idea of our approach is to estimate the model's nonlinear component by means of regression splines after projecting the variables in the nonlinear term to a single covariate. 
    \item The primary findings of the analyses conducted on the NHANES are:
    \begin{enumerate}
        \item The proposed single-index model is shown to outperform both the global Fréchet model and a competing, more general, partially linear Fr\'echet regression model in terms of the adjusted Fréchet $R^2$ measure. 
        \item Interpretations are provided for the effects of ethnicity and other interesting variables on distributional physical activity profiles. These novel analyses can provide new insights into how physical activity varies among the various US sub-populations.
        \item New physical activity phenotypes are constructed corresponding to individuals who do more or less exercise than is predicted by the model using the distributional representation.  These analyses are new and help to examine how well individuals adhere to the recommended physical activity guidelines.
        \end{enumerate}
\end{itemize}
The code for reproducing the results presented using the methods proposed in this paper are publicly available on GitHub at  \url{https://github.com/aghosal89/FSI_NHANES_Application}.

\section{Motivating example: data on wearable accelerometer devices from NHANES 2011-2014}
\label{sec:nhanes}

The NHANES aims to provide a broad range of
descriptive health and nutrition statistics for the non-institutionalized civilian US population \cite{johnson2014national}.
Data collection consists of an interview and an examination. The interview gathers personal
demographic, health, and nutrition information; the examination includes physical measurements such as blood pressure, a dental examination, and the collection of blood and urine specimens for laboratory
testing. Additionally, participants were asked to wear a physical activity monitor, starting on the day of
their exam, and to keep wearing this device all day and night for seven full days (midnight to midnight) and remove it on the morning of the 8th day. The device used was the ActiGraph GT3X+ (ActiGraph of
Pensacola, FL). Data from the NHANES cohorts 2011--2014 were used for the analyses in this paper \cite{johnson2014national}. 

Physical activity signals were pre-processed by staff from the National Center for Health Statistics (NCHS) to determine signal patterns that were unlikely to result from human movement. Acceleration measurements were then summarized at the minute level using Monitor-Independent Summary (MIMS) units, an open-source, device-independent universal summary metric \cite{john2019open}. In order to further increase the reliability of the analysis, we use the following filter criteria strategy extracted from \cite{Wsmirnova2019predictive} in order to remove participants with poor quality in their accelerometry data. Those participants who (i) had fewer than three days of data with at least 10 hours of estimated wear time or (ii) had non-wear periods, identified as intervals with at least 60 consecutive minutes of zero activity counts and at most 2 minutes with counts between 0 and 5 were deemed by NHANES to have poor quality and hence were removed. These protocol instructions were adapted from high-level accelerometer research (see, for example, \cite{troiano2008physical}).

\subsection{Quantile function representation of physical activity profiles}
\label{ss:quantile_representations}
A novel representation of the resulting data is herein adopted that extends previous compositional metrics to a functional setting \cite{10.1093/jrsssc/qlad007}, aimed at overcoming their dependency on certain physical activity intensity thresholds. This approach also overcomes some previously known limitations of more traditional approaches. Let $i\in \{1,2,\dots,n\}$ be the index for participants, where $n$ is the total number of participants in the study. For the $i$-th participant, let $M_i$ indicate the number of days (including partial days) for which accelerometer records are available and $n_i$ be the number of observations recorded in the form of pairs $\left(m_{ij},A_{ij}\right)$, $j=1,\dots,n_i$.  Here, the $m_{ij}$ are a sequence of time points in the interval $\left[0,M_i\right]$ in which the accelerometer records activity information and $A_{ij}$ is the measurement of the accelerometer at time $m_{ij}$. No data are available during non-wear periods.  

\begin{figure}  
\includegraphics[scale=0.75]{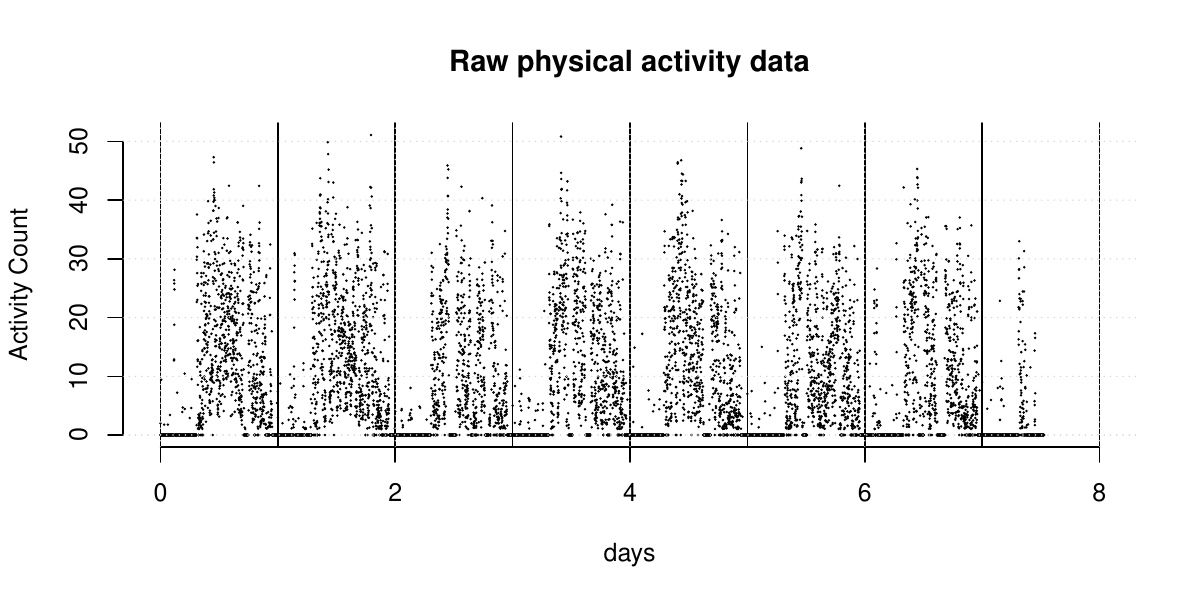}
\caption{\footnotesize The plot of physical activity time series $A_{ij}$ of one representative participant (one chosen $i$) in the NHANES 2011-2014 study monitored during 8 days are plotted over the observed time intervals $m_{ij}$, when the physical activity measurements are counted as described in Section \ref{ss:quantile_representations}.}
\label{fig:raw}
\end{figure}

\begin{figure}
\includegraphics[scale=.72]{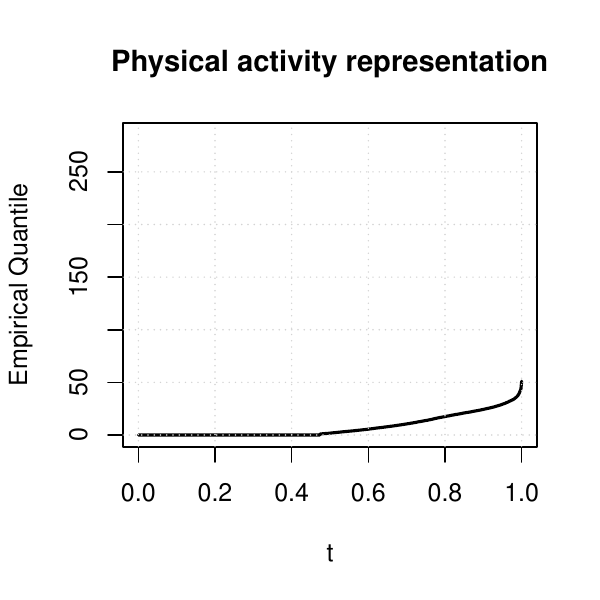}
\includegraphics[scale=.72]{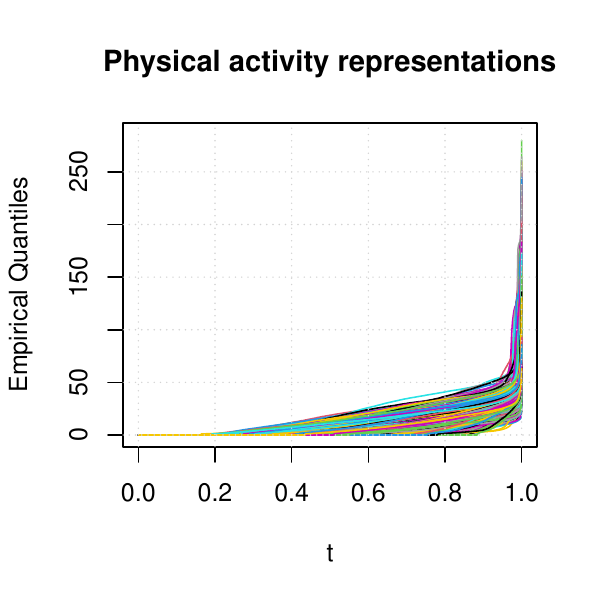}
\caption{\footnotesize (Left) The empirical quantile representation $\hat{Q}_i$, of the activity profile of the participant (chosen $i$) described in Figure~\ref{fig:raw} above, also described in the Section \ref{ss:quantile_representations}. (Right) The estimated empirical quantile functions of physical activity profiles 
for all of the 4616 participants in the study.} 
\label{fig:raw2}
\end{figure}

In this paper, each individual's accelerometer measurements $\{A_{ij}\}_{j=1}^{n_i}$, $i=1,\dots,n$, are studied without regard to their ordering. They are thus characterized by the empirical quantile function, $Y_i(t)=\hat{Q}_i(t)$, for $t \in [0,1]$, which will be used as the response in the regression model. Here, $\hat{Q}_i(t)= \inf\{a\in \mathbb{R}: \hat{F}_{i}(a)\geq t\} $ is the generalized inverse of $\hat{F}_{i}\left(a\right)=\frac{1}{n_i} \sum_{j=1}^{n_i} 1\{A_{ij}  \leq a \}$, $a \in \mathbb{R}$, the empirical cumulative distribution function for the physical activity values for the $i$-th individual. In order to illustrate clearly the difficulty of analyzing raw physical activity data from participants who are monitored during different periods and in different experimental conditions, Figure \ref{fig:raw} shows the plot of observed $A_{ij}$ against $m_{ij}$ for an arbitrary participant in our study. In Figure \ref{fig:raw2}, the left panel shows the empirical quantile functional representation of the physical activity measurements of the participant whose raw measurements are shown in Figure \ref{fig:raw}, while the right panel shows the empirical quantile functions of all participants after transforming the raw time series physical activity data into distributions of the physical activity. Quantile function representations of the physical activity trajectories overcome the problems of the traditional summary metrics of physical activity when the raw time series have different lengths. In addition, the new representation uses all accelerometer intensities (over a continuum) to construct the new physical activity functional profile, generalizing traditional metrics of physical activity that summarize the information in a compositional vector.


The precise relationship between the distributional representation and traditional compositional metrics can be understood as follows. Given a subject's cumulative distribution function (CDF) of physical activity, $\hat{F}_i(a)$, a compositional metric is formed by creating bins $[a_{j-1}, a_j]$ for $a_0 < a_1 < \ldots < a_M$. This process results in an $M$-dimensional compositional vector $\mathbf{U}_i$, where each component is defined as

\[
U_{ij} = \hat{F}_i(a_j) - \hat{F}_i(a_{j-1}).
\]

\noindent Thus, any desired compositional information can always be derived from the distribution, and the distributional representation can be thought of as a continuous extension of the compositional approach. Most importantly, when physical activity distributions are the responses in a regression model, distributional predictions produced by the regression fit  can always provide \emph{any} desired compositional information regardless of the choice of bins or thresholds $a_j$, \(j = 0, 1, \ldots, M\).

Hence, from a practical perspective, there are two main advantages to choosing the distributional representation.  First, it avoids collapsing the information of physical activity into intervals, preventing potential loss of information. Second, it eliminates the need to define thresholds that, in the context of physical activity, depend on population and physical characteristics such as age, sex, and BMI. This removes arbitrariness and subjectivity from the analysis.

Another key distinction between this work and others that have used distributional representations is that, while these previous papers have focused only on positive observations of physical activity counts \cite{ghosal2023distributional,lin2023causal}, the current work also incorporates periods of recorded inactivity. Inactivity behavior is an important component of human physical activity profiles, and the proposed methodological approach allows for the integration of both inactivity and activity behaviors comprehensively.

\subsection{Details of covariates}
\label{ss:covariates}

In addition to the accelerometer data, the covariates used in the model include sociodemographic, dietary, and clinical variables such as age, Body Mass Index (BMI), and Healthy Eating Index (HEI), 
along with the categorical variables Ethnicity and Sex and their interaction. HEI is a continuous score reflecting the overall diet quality of each participant. The ethnicity variable reported the racial origin of the participants divided into the following 6 categories: Mexican American, Other Hispanic, Non-Hispanic White, Non-Hispanic Black, Non-Hispanic Asian, and Other races, including Multi-racial.  The age range of the participants in the analysis was 20 to 80 years. The BMI ($\text{Kg/m}^2$) was restricted to the range $18.5 - 40$ to study individuals ranging from healthy to highly overweight/obese. Under these restrictions, $n = 4616$ individuals were chosen for the analysis.  Although not included as a predictor in the models analyzed in Section~\ref{sec:results}, the Total Activity Count (TAC) is a widely used summary metric for accelerometer data that represents the average activity count over the complete series of physical activity measurements recorded by the device. In the data cohort used for analysis, each element of the  sequence \(\{A_{ij}\}_{j=1}^{n_i}\) is measured in MIMS units, unlike NHANES 2003-2006, which uses Actigraph counts. For consistency, since the TAC variable in both datasets represents the average count units of the monitor, this metric  is termed Total Activity Count (TAC) in both settings. More specifically, in the notation of Section \ref{ss:quantile_representations}, the TAC of the $i$-th participant is $TAC_{i}=\frac{1}{n_i}\sum_{j=1}^{n_i}A_{ij}$. In this sense, TAC represents a scalar summary of the distributional representation used in this paper. Section demonstrates how the proposed model, which relates covariates to the full physical activity distrbution, can subsequently be used to interpret relationships between covariates and summaries of interest such as TAC. 
Univariate summaries of each of these covariates, stratified by sex, are displayed in Table~\ref{tab:variables_in_analysis}.

\begin{table}[t]
\centering
\caption{\footnotesize Summaries of the covariates Age, BMI (Body Mass Index), HEI (Healthy Eating Index), TAC (Total Activity Count), and Ethnicity, stratified by Sex. In the first column, the levels of the categorical variable Ethnicity are separated from the numerical covariates Age, BMI, HEI, and TAC. In the third and fourth columns, the first four rows present the means and, in brackets, the standard deviations of the continuous variables (Age, BMI, HEI, and TAC) for men and women respectively. Rows 5 - 10 in the same columns represent the percentage breakdown of the sub-populations of men and women into their respective ethnicities. The description of the covariates are found in Section \ref{ss:covariates}.
\label{tab:variables_in_analysis}}
\smallskip
\begin{tabular}{|l|r|c|c|}
  \hline
 & Covariates & Men & Women
 \\ 
  \hline
& Age & 47.45 (16.45) & 48.082 (16.50) \\ 
 Numeric & Body Mass Index & 28.72 (5.73) & 29.18 (7.41) \\ 
 Variables & Healthy Eating Index & 53.013 (14.13) & 56.63 (14.75) \\ 
 & Total Activity Count & 9.8 (3.1) & 9.7 (2.6) \\ \hline &
  Mexican American & 8.55 \% & 6.42 \% \\ 
 & Other Hispanic & 5.42 \% & 5.28 \% \\ 
 Ethnicities & Non-Hispanic White & 70.65 \% & 72.3 \% \\ 
 & Non-Hispanic Black & 8.82 \% & 10.29 \% \\ 
 & Non-Hispanic Asian & 3.77 \% & 3.37 \% \\ 
 & Other Races Including Multi-racial & 2.79 \% & 2.35 \%\\ \hline
\end{tabular}
\end{table}
This paper aims to create a parsimonious and straightforward regression model to interpret the several central aspects of energetic expenditure captured by the Age and BMI variables that are expected to behave in a nonlinear way with the response. At the same time, it is of interest to assess the effect of diet on physical exercise. It has been observed that sex and ethnicity differences in the US population tend to interact concerning physical activity. For instance, women tend to be physically less active than men within some ethnic groups (Black, While, Asian, Other races, including Multi-racial) \cite{troiano2008physical},  while among the Mexican American and Other Hispanic ethnicities, men and women show very similar physical activity levels 
 \cite{ortiz2010sociodemographic} and adjusted for demographic factors \cite{medina2013physical}. Hence, an interaction between sex and ethnicity was included to obtain reliable population-based conclusions about the relationships between these covariates and physical activity. The inherent sampling design of the NHANES provides essential advantages to obtaining reliable population measurements that cannot be guaranteed with observational cohorts such as the UK-Biobank due to selection bias. In order to properly exploit this advantage, however, the survey design must be taken into account in the estimation procedure, as described in Section~\ref{ss:est}.  

\section{The Partially Linear Fréchet Single Index Regression Model}
\label{sec:algorithm}

Let $Y_i$ be the empirical quantile function of daily activity levels corresponding to the $i$-th participant. In what follows, the regression is built by directly modeling the pointwise mean function of $Y_i(t)$ on the covariates, $t \in [0,1]$. Using the quantile function to characterize the physical activity distribution can be explained as follows. First, a density or cumulative distribution function (cdf) representation that ignores inactivity time is inappropriate because the distributions represented by the $Y_i$ are a mixture of a mass at zero and a continuous distribution for positive values. There are other practical reasons to prefer quantile functions.  For instance, quantile functions are less constrained than cdfs or density functions. For example, one may add two quantile functions or multiply one by any positive constant to produce another, but not so for cdfs or densities.  This is crucial for predictive modeling, since applying a post hoc adjustment to a model prediction to obtain a valid distributional representation can affect both interpretation and introduce distortions that are more prominent for cdfs or densities than for quantile functions \cite{petersen2021wasserstein}.  Second, modeling the mean quantile function is directly related to \textit{optimal transport} through the \textit{Wasserstein metric}, offering a natural framework for understanding biological phenomena \cite{zhang2011functional,villani2009optimal, peyre2019computational, panaretos2019statistical}. 


Briefly, if $\mu$ and $\nu$ are two suitable measures on $\mathcal{R}$ with finite second moment, and if $Q_\mu$ and $Q_\nu$ are their corresponding quantile functions, then $d_{W_2}(\mu,\nu),$ the Wasserstein distance between $\mu$ and $\nu$, is known to be equivalent to the $L^2$ distance between $Q_\mu$ and $Q_\nu$, that is,
\begin{equation}
    \label{eq:wassL2equiv}
    d_{W_2}(\mu,\nu) = \left[\int_0^1 (Q_\mu(t) - Q_\nu(t))^2\mathrm{d}t\right]^{1/2}.
\end{equation}
As a consequence, under this metric, the Fréchet mean \cite{frec:48} measure of a random measure is characterized by the pointwise mean of the corresponding random quantile process.  Hence, by proposing a regression model for the random quantile function $Y_i,$ one is implicitly constructing a model for the conditional (Wasserstein-)Fréchet mean of the underlying random physical activity distribution measure \cite{petersen2021wasserstein}.

The partially linear Fréchet single index (PL-FSI) model is formally defined as follows. 
Let $\boldsymbol{X_{i}}\in \mathcal{R}^p$ denote the $p$-dimensional covariate vector in the single index part of the model, while $\boldsymbol{Z_i} \in \mathcal{R}^q$ is the covariate vector considered for the linear part. The PL-FSI model is
\begin{equation}
E(Y_i(t)|\boldsymbol{X}_i,\boldsymbol{Z}_i)=\alpha(t) + \boldsymbol{\beta}(t)^{T}\boldsymbol{Z}_i+g(\boldsymbol{\theta}_0^{T}\boldsymbol{X}_i,t),\quad  t\in [0,1],
\label{eq:wass_regression_model}
\end{equation} 
where the vector $\boldsymbol{\theta_0} \in \mathcal{R}^p$, intercept function $\alpha$, coefficient function $\boldsymbol{\beta}$ and link function $g$ are the unknown parameters.  

\subsection{Model Estimation}
\label{ss:est}

For estimating the parameter $\boldsymbol{\theta}_0$ and the identifiability of the PL-FSI model \cite{lin2007identifiability}, define the parameter space $$\Theta_p=\{\boldsymbol{\theta}\in \mathcal{R}^p: \norm{\boldsymbol{\theta}}_E=1, \text{ first non-zero element being strictly positive} \}$$  
where $\norm{\cdot}_E$ is the Euclidean norm. To facilitate estimation of the smooth bivariate function $g,$ the expansion
\begin{equation}
g(u,t) \approx \sum_{k=1}^{K+s}{\gamma}_k(t)\phi_k(u)
\label{eq:g_ui_t spline approx}
\end{equation}
will be used, where $\{\phi_k\}_{k=1}^{K+s}$ is a B-spline basis of order $s$ on $K$ interior knots, and ${\gamma}_k(t)$ are the coefficients of the basis as a function of $t$.

In practice, the tuning parameter $K$ can be chosen by evaluating candidate values using a goodness-of-fit measure, such as the coefficient of determination $R^2$ or related measures, while $s = 4$ is usually chosen, corresponding to cubic splines.
The approximation to the PL-FSI model \eqref{eq:wass_regression_model} can thus be written as
\begin{equation}
E(Y_i(t)|\boldsymbol{X}_i,\boldsymbol{Z}_i) \approx \alpha(t) + \boldsymbol{\beta}(t)^{T}\boldsymbol{Z}_i+\boldsymbol{\gamma}(t)^{T}\boldsymbol{U}_i(\boldsymbol{\theta}_0),\quad  t\in [0,1],
\label{eq:wass_regression_basis}
\end{equation} 
where $\boldsymbol{\gamma}(t) = (\gamma_1(t),\ldots,\gamma_{K + s}(t))^{T}$ and, for any $\boldsymbol{\theta} \in \Theta_p,$ $\boldsymbol{U}_i(\boldsymbol{\theta}) = (\phi_1(\boldsymbol{\theta}^{T}\boldsymbol{X}_i),\ldots,\phi_{K+s}(\boldsymbol{\theta}^{T}\boldsymbol{X}_i))^{T}.$

As \eqref{eq:wass_regression_basis} exhibits a linear form for each fixed value of $\boldsymbol{\theta}$, a semi-parametric least-squares approach can be utilized for estimation. Due to the complex survey design of the NHANES database, a weighted least squares criterion is needed in order to perform inference correctly and obtain reliable results \cite{lumley2010svy}. Assume that a sample $\mathcal{D}=\{\left(Y_i, \boldsymbol{X}_i,\boldsymbol{Z}_i \right): i \in S\}$ is available, where $Y_i$ is a response variable, and $\boldsymbol{X}_i, \boldsymbol{Z}_i$ are vectors of covariates taking values in a finite-dimensional space. The index set $S$ represents a sample of $n$ units from a finite population. 

To account for sampling, each individual \( i \in S \) is associated with a positive weight \( w_i \) derived from an experimental design such as multistage random sampling. In the particular case of NHANES, the survey weights \( w_i \) are specified by the Centers for Disease Control and Prevention (CDC)\footnote{\url{https://wwwn.cdc.gov/nchs/nhanes/search/datapage.aspx?Component=Demographics\&CycleBeginYear=2011}} and are primarily used to mitigate selection bias, as explained in the CDC guidelines\footnote{\url{https://wwwn.cdc.gov/nchs/nhanes/tutorials/weighting.aspx}}, to obtain reliable conclusions about the US population.  In the analyses conducted in this paper, these weights were taken to be the inverse of the probability \( \pi_i > 0 \) of being selected into the sample, i.e., \( w_i = \frac{1}{\pi_i} \) \cite{kish1965survey, lumley2004analysisn}. 
These weights are used to construct an estimator of Horvitz-Thompson type \cite{horvitz1952generalization, https://doi.org/10.1111/j.1467-985X.2006.00426.x} by constructing a weighted least squares criterion. 

The full procedure can be broken down into two steps. In the first step, for any $\boldsymbol{\theta} \in \Theta_p$ and any $t \in [0,1],$ one computes weighted least squares estimates
\begin{equation}
\left(\hat{\alpha}_{\boldsymbol{\theta}}(t),\hat{\boldsymbol{\beta}}_{\boldsymbol{\theta}}(t), \hat{\boldsymbol{\gamma}}_{\boldsymbol{\theta}}(t)\right)=\underset{a \in \mathcal{R},\boldsymbol{b}\in \mathcal{R}^q,\boldsymbol{c}\in 
\mathcal{R}^{K+s}}{\operatorname{\textrm{argmin}}}\, \sum_{i=1}^{n} w_i\left[Y_{i}(t)-a-\boldsymbol{b}^{T} \boldsymbol{z}_{i}-\boldsymbol{c}^{T} \boldsymbol{U}_{i}(\boldsymbol{\theta})\right]^{2}.
\label{eq:reg_param_est}
\end{equation}
These estimates lead to initial fitted quantile functions
\begin{equation}
Y_{i}^{*}(\boldsymbol{\theta}, t)=\hat{\alpha}_{\boldsymbol{\theta}}(t)+\hat{\boldsymbol{\beta}}_{\boldsymbol{\theta}}^{T}(t) \boldsymbol{Z}_{i}+\hat{\boldsymbol{\gamma}}_{\boldsymbol{\theta}}^{T}(t) \boldsymbol{U}_{i}(\boldsymbol{\theta}),\quad t \in [0,1].
\label{eq:Yhat_theta_t}
\end{equation}
However, as a function of $t$, $Y_i^*(\boldsymbol{\theta},t)$ may not be monotonically increasing and hence is not a proper quantile function. The typical solution for this is to project $Y_i^*(\boldsymbol{\theta},t)$, in the $L^2[0,1]$ sense, onto the nearest monotonic function \cite{petersen2021wasserstein}, yielding a valid quantile function $\hat{Y}_i(\boldsymbol{\theta},t).$ For more details, the reader is referred to Algorithm~2 in the supplementary material of ~\cite{petersen2021wasserstein}. This algorithm produces a non-decreasing quantile function because the space of quantile functions is closed and convex, ensuring that the projection is unique. The algorithm used is a quadratic program solver that effectively adjusts the estimated quantile function to enforce monotonicity, thereby guaranteeing that the resulting function is non-decreasing.

Once these initial quantities are formed for any $\boldsymbol{\theta}$ and $t$, an estimate $\hat{\boldsymbol{\theta}}_0$ can be computed.  As justified in \cite{10.1214/23-EJS21202}, one can use a generalized version of the residual sums of squares to obtain the estimate. In the current context, the survey-weighted criterion
\begin{equation}
W_{n}(\boldsymbol{\theta})=
\sum_{i=1}^{n} w_i \int_{0}^{1}\left\{Y_{i}(t)-\hat{Y}_{i}(\boldsymbol{\theta}, t)\right\}^{2} d t
\label{eq:W_n_function}
\end{equation}
is proposed, and constitutes a weighted average of the squared $L^2$ norms of the quantile residuals (or, equivalently, of the squared Wasserstein distances between observed and fitted physical activity distributions).  Then the estimated parameter is
\begin{equation}
\hat{\boldsymbol{\theta}}=\underset{\boldsymbol{\theta}\in \Theta_p}{\operatorname{\textrm{argmin}}}\, W_{n}(\boldsymbol{\theta}).
\label{eq:theta_estimate}
\end{equation}

From this estimate of the index parameter, given any covariate pair $(\boldsymbol{z},\boldsymbol{x}),$ the conditional Wasserstein-Fréchet mean quantile function can be estimated as follows.  First, the basis functions are evaluated at the relevant input by computing $\hat{\boldsymbol{u}} = (\phi_1(\hat{\boldsymbol{\theta}}^{T}\boldsymbol{x}),\ldots,\phi_{K+s}(\hat{\boldsymbol{\theta}}^{T}\boldsymbol{x}))^{T}.$ Then, \eqref{eq:reg_param_est} is computed at the specified $\hat{\boldsymbol{\theta}}$ and, as in \eqref{eq:Yhat_theta_t}, the preliminary estimate
\begin{equation}
Y^*(t; \boldsymbol{z},\boldsymbol{x}) = \hat{\alpha}_{\hat{\boldsymbol{\theta}}}(t) + \hat{\boldsymbol{\beta}}^{T}_{\hat{\boldsymbol{\theta}}}(t)\boldsymbol{z} + \hat{\boldsymbol{\gamma}}_{\hat{\boldsymbol{\theta}}}^{T}(t)\hat{\boldsymbol{u}}.
\label{eq:Y_est_notL2}
\end{equation}
is constructed.  Finally, the estimated quantile function $\hat{Y}(t;\boldsymbol{z},\boldsymbol{x})$ is obtained by projecting (if necessary), in the $L^2$ sense, $Y^*(t;\boldsymbol{z},\boldsymbol{x})$ onto the space of quantile functions, meaning the nearest monotonically increasing function. In particular, for any set of observed covariates $(\boldsymbol{Z_i}, \boldsymbol{X_i}),$ fitted values $\hat{Y}_i(t) = \hat{Y}(t;\boldsymbol{Z_i}, \boldsymbol{X_i})$ are obtained.

Importantly, the proposed estimation method differs from other additive functional regression models in the literature, such as those presented by \cite{mclean2014functional}, which represent quantile functions using a basis. A basis representation for the quantile response objects is not required for the several reasons: first, it is not necessary for implementation of the proposed estimation algorithm, and its use would not simplify computation; second, it does not provide additional insights into the biological problem or aid in interpretation; and, third,
the zero pattern in quantile functions does not lend itself well to such representations in general, except for special bases like B-splines. Furthermore, another alternative used in the nonparametric regression literature is to employ kernel smoothing to estimate the nonparametric part. However, local smoothers are more difficult to incorporate with linear effect estimation compared to global smoothers like splines.

\subsection{Model Inference: Global Confidence bands via survey boostraping}
\label{ss:boot}



In order to quantify uncertainty related to estimates \( \hat{\beta}(t) \) in the linear part of the model, a survey bootstrapping methodology was used with the goal of deriving global confidence bands for each linear predictor. From a practical point of view, deriving global confidence bands for functional predictors provides more insights into the statistical significance of the variables, overcoming the limitations of pointwise confidence bands that suffer from multiple comparisons.

To achieve this, the survey bootstrap methodology introduced for multistage designs in \cite{doi:10.1177/096228029600500305} was used. 
The high-level idea is to introduce a multiplier bootstrap methodology. For each participant \( i \in \{1, \dots, n\} \) and each bootstrap replicate \( b \in \{1, \dots, B_s\} \), a multiplier weight \( \omega_{i}^{(b)} \) is chosen. The values of the random multipliers \( \omega_{i}^{(b)} \) depend on certain tuning parameters, such as the number of observations drawn from each stratum (which depends on the geographical Primary Sampling Units (PSUs) area) and the rate of bootstrap with replacement. Technical details about the multiplier calculations can be found in \cite{doi:10.1177/096228029600500305}, where closed-form expressions for the weights are provided in some special cases.

Let \( r = 1, \dots, q \) index the different components in the linear term, \( i = 1, 2, \ldots, n \) be the index for the participants in the study, and \( b = 1, 2, \ldots, B_s \) be the index for bootstrap replicates. The bootstrap survey weights are \(w_i^{(b)} = \omega_i^{(b)}w_i\), where $w_i$ are the original survey weights.  Let \( w^{(b)} = \left( w_1^{(b)}, w_2^{(b)}, \ldots, w_n^{(b)} \right)^\top \) be the \( b \)-th bootstrap sampling weights for all participants, whereas \( w = \left( w_1, w_2, \ldots, w_n \right)^\top \) are the sampling weights assigned by the NHANES for the individuals under study.

Let \( \hat{\beta}_r^{(b)}(t) \) be the estimate of the $r$-th regression parameter \( \beta_r(t) \) for the \( b \)-th simulation and the \( t \)-th quantile, using the same estimation process as described in Section~\ref{ss:est}, replacing the survey weights \( w_i \) with \( w_i^{(b)} \). 
 Define the pointwise boostrap means and variances as
\[
\bar{\hat{\beta}}_r(t) = \frac{1}{B_s} \sum_{b=1}^{B_s} \hat{\beta}_r^{(b)}(t), \quad s_{r,\hat{\beta}}^2(t) = \frac{1}{B_s - 1} \sum_{b=1}^{B_s} \left( \hat{\beta}_r^{(b)}(t) - \bar{\hat{\beta}}_r(t) \right)^2,
\]

then calculate
\[
u_{r,b} = \underset{t \in (0,1)}{\sup} \, \frac{\left| \hat{\beta}_r^{(b)}(t) - \hat{\beta}_r(t) \right|}{s_{r,\hat{\beta}}(t)}, \quad  b = 1, 2, \ldots, B_s; \ r = 1, 2, \ldots, q.
\]
The upper \( 95\% \) quantile of \( \{ u_{r,1}, u_{r,2}, \ldots, u_{r,B_s} \} \) is denoted by \( q_{0.05}^{(r)} \). Hence, the \( 95\% \) bootstrap confidence interval for the parameter \( \beta_r(t) \) is
\[
\left( \hat{\beta}_r(t) - q_{0.05}^{(r)} s_{r,\hat{\beta}}(t), \quad \hat{\beta}_r(t) + q_{0.05}^{(r)} s_{r,\hat{\beta}}(t) \right).
\]



\subsection{Computational Details}
\label{subsec:comp_details}

Details regarding the implementation of the proposed PL-FSI model and its estimation for the NHANES database will now be provided. In the models implemented below in Section~\ref{sec:results}, the nonlinear covariate $\boldsymbol{X_{i}}$ for the $i$-th individual consists of their BMI and Age, so the dimension for this component is $q = 2,$ i.e., $\boldsymbol{\theta}_0 \in \Theta_2$. For the spline basis in \eqref{eq:g_ui_t spline approx}, computations were performed using the \texttt{dbs} function in the package \texttt{splines2} \cite{wang_yan_2021}. Knot placement was determined internally by the default option of the \texttt{dbs} function and varied with the value of $\boldsymbol{\theta}$.  Specifically, for any $K$, equally spaced values $r_k,$ $k = 1,\ldots,K,$ were computed, where $r_0 = 0 < r_1 < \cdots < r_K < r_{K + 1} = 1;$ the $k$-th interior knot was then taken as the $r_k$-th empirical quantile of the values $\{\boldsymbol{\theta}^{T}X_i;\, i = 1,\ldots,n\}.$
In the experiments conducted, $s=4$ and $K=5$ were used, so the number of spline regression parameters was $K + s = 9$. The choice $K = 5$ was made based on a preliminary investigation of the PL-FSI model fit for potential values $K = 4,\ldots,10$.  The adjusted Fr\'echet $R^2$ criterion, defined below in Section~\ref{sec:results}, showed a jump from $K = 4$ to $K = 5$, followed by stable values for larger $K$. The covariates in the linear component $\boldsymbol{Z}_i$ consist of HEI (continuous) and indicator variables for Sex and Ethnicity, as well as the interaction between these. 

The estimates of parameters in \eqref{eq:reg_param_est} can be efficiently computed as a weighted least squares problem for any fixed $\boldsymbol{\theta}$ and $t \in [0,1]$. However, this can only be done in practice for a finite ordered grid of values $t \in T_m = \{t_1,\ldots,t_m\} \subset[0,1]$. 
These initial survey-weighted least squares computations were done using R package \texttt{survey} \cite{lumley2004analysisn,lumley2010svy,lumley2020svy}, which allows for the introduction of splines into the regression model while simultaneously computing and incorporating the survey weights $w_i$.

For any given $\boldsymbol{\theta}$ and grid point $t$, computation of \eqref{eq:Yhat_theta_t} is straightforward. To execute the projection step, observe that monotonicity can only be achieved in the discrete sense in dependence on the chosen grid $T_m$.  We refer to \cite{petersen2021wasserstein} for a simple description of this projection algorithm, which can be done using any basic quadratic program solver.  Consequently, for a given $\boldsymbol{\theta},$ \eqref{eq:W_n_function} is approximated by numerical integration. Finally, to perform the optimization in \eqref{eq:theta_estimate}, the function $\texttt{optim}$ in R was used with the L-BFGS-B algorithm by repeatedly performing the above steps to evaluate $W_n(\boldsymbol{\theta})$ for different values of $\boldsymbol{\theta}$ across iterations. To deal with the possibility of local minima, four different starting values (taken to be equally spaced in their angular representation) in $\Theta_2$ were used for this optimization step, yielding four (possibly not unique) candidate estimators at convergence. The final estimator was taken to be the candidate yielding the smallest value of $W_n.$

\section{Experimental Results}
\label{sec:results}

This section explores the PL-FSI model when it is fitted to the NHANES database, with physical activity distributions as the response functions. First, the PL-FSI model is compared to two competing models, the global Fr\'echet regression model and an alternative semi-parametric Fr\'echet regression model.  This latter model replaces the single index term in \eqref{eq:wass_regression_model} with two separate additive terms $g_1(X_{i1}) + g_2(X_{i2})$, and is termed a Partially Linear (Additive) Fr\'echet (PLF) model.  The PLF model is a natural candidate for including nonlinear effects for Age and BMI.  While it does not impose a single index structure, it also does not account for any interaction between these two covariates, which is a critical biological feature. The PLF model was similarly fitted using two separate cublic spline terms, each with $K = 5$ knots.

After these initial model comparisons, the fitted linear and nonlinear components of the PL-FSI model are interpreted in turn in order to identify the implied associations with the various covariates. Beginning with the predictors in the linear component of the model, differences between diverse subpopulations corresponding to ethnicity and sex are investigated.  From there, the combined associations of BMI and age are elucidated through the spline fit of the nonlinear model component.  Throughout, due to the use of the quantile representation of physical activity, comparisons are made across the range of physical activity intensities, rather than at the mean intensity or some pre-specified selection of quantiles as is the common practice.

\subsection{Comparison of Competing Models}
\label{ss:modelComp}

To provide an even comparison, the global Fréchet (GF) model was slightly modify by introducing the specific survey weights in the estimation criterion. The covariates used were the same in each of the GF, PL-FSI, and PLF models, with the only difference being that the latter two included the BMI and Age in their nonlinear single index and nonlinear additive components, respectively. Hence, the global Fréchet model can be considered as a special case of both the PL-FSI and PLF models, in which all covariates are included in the linear component.  To facilitate interpretation, all numerical covariates were centered and scaled prior to fitting the models. 




To quantify differences in the quality of model fits, the capacity of the models to explain differences in physical activity distributions across individuals can be evaluated using the survey-weighted Fréchet $R^{2}$
\begin{equation}
    R_{\oplus}^{2}=  1-\frac{\sum_{i=1}^{n} w_i \int_0^1 (Y_i(t)-\hat{Y}_i(t))^2 \mathrm{d}t}{ \sum_{i=1}^{n} w_i  \int_0^1 (Y_i(t)-\overline{Y}(t))^2 \mathrm{d}t}
    \label{eq:fr_rsq}
\end{equation} 
where $\overline{Y}(t)= \left(\sum_{i = 1}^n w_i \right)^{-1}\sum_{i = 1}^n w_i Y_i(t)$ is the weighted sample Wasserstein-Fréchet  mean of the observed physical activity distributions and $w_i$ is the survey weight corresponding to $i$-th observation.  Since the weights \( w_i \) are positive, $\overline{Y}$ is a convex combination of the individual quantile functions \( Y_i \), ensuring that it is non-decreasing and thus a valid quantile function.

Because the models have differing levels of complexity even with the same set of predictors, $R_\oplus^2$ does not provide a fair comparison.  Hence, the adjusted Fréchet $R^2$, denoted as $\bar{R}_{\oplus}^{2}$, was also computed.  With $n$ being the number of observations and $q'$ the number of regression parameters included in the model being considered,
\begin{equation}
\bar{R}_{\oplus}^{2}={R}_{\oplus}^{2}-\left(1-{R}_{\oplus}^{2}\right) \frac{q'}{n-q'-1}
\label{eq:adj_fr_rsq}
\end{equation} 
enables a fair comparison of the quality of model fit that adjusts for the different complexity of each model.

The PL-FSI model had a higher $\bar{R}_\oplus^2$ value of 0.146, which is 24\% higher relative to the 0.118 value obtained by the global Fréchet model. This suggests that although the predictive capacity of both models is moderate, the additional parameters introduced by the single index and spline representation improved the variance explained. The overall moderate statistical associations are not surprising given the multitude of factors that can influence high-dimensional physical activity distributions, many of which are not included in our model or the NHANES database.  Nonetheless, we acknowledge this limitation and emphasize the interpretative value of the detected associations in the next sections with functional coefficients despite the moderate \( R^2 \) values.

\begin{figure}[t]
    \centering
    \includegraphics[width=.48\textwidth]{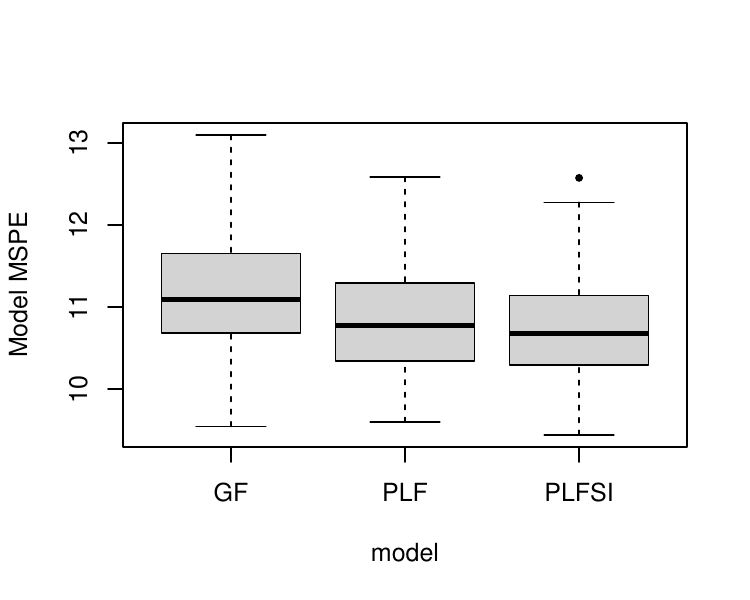} 
    \includegraphics[width=.48\textwidth]{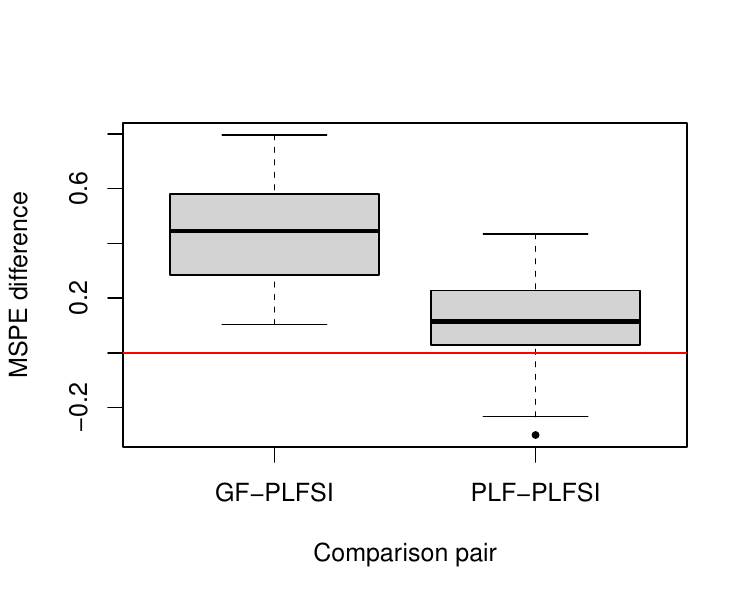} 
    \caption{\footnotesize (Left) Out-of-sample mean square prediction errors across 40 independent data splits for the Global Fréchet (GF), Partially Linear Fr\'echet (PLF), and Partially Linear Fr\'echet Single Index (PL-FSI) models. (Right) Differences and out-of-sample mean square prediction errors of GF and PLF models, respectively, relative to the PL-FSI model, with horizontal red line at 0 for reference.}
    \label{fig:mspe_comparison}
\end{figure}

To provide a more reliable predictive evaluation between the three models, cross-validation was conducted using 40 independent data splits (90\% training and 10\% testing), and the empirical performance across three models was compared in terms of the mean square prediction error on the test sets, quantified as the average squared Wasserstein distance between predicted and observed physical activity quantile functions. Empirically, the GF and PLF models are both outperformed in terms of out-of-sample prediction by the proposed PL-FSI model. Figure \ref{fig:mspe_comparison} visualizes the error comparison of the three models considered, each with different levels of complexity and structural properties. The PL-FSI) model strikes a balance between the restrictive global Fréchet regression model and the more flexible PLF model, while also accounting for the age-BMI interaction through the single index—a feature that neither of the other models accommodates.

\subsection{Interpretation of the PL-FSI Model Fit}
\label{subsec:interpret}

A salient advantage of the PL-FSI model is its diverse model components that incorporate linear, categorical, and nonlinear effects. These components are harmonized into the final regression model through an additive regression structure. It is important to stress that, while the model is accurately described as linear in the way that it models covariate associations at each $t \in [0,1]$, the nature of the model is entirely nonparametric in its treatment of the variations across $t$ of both the coefficient functions and, by extension, the fitted quantile functions of physical activity.

A important practical issue arises from the fact that, due to the final projection step in obtaining fitted quantile functions, the interpretability of coefficient function estimates from the linear component could be questionable. In the setting of global Fréchet regression, Lemma 2 in \cite{petersen2021wasserstein} demonstrated that, as \( n \) grows to infinity, the probability that the projection step is necessary for computing a prediction for any of the $n$ data points converges to zero. Given that the analyzed cohort consists of over 4,000 individuals, the effect of projection is expected to be minor.

In Sections~\ref{subsec:HEI_interpret}--\ref{subsec:nonlinear_interpret}, the estimated regression parameters in \eqref{eq:Y_est_notL2} will be interpreted and discussed.  Due to the boundary effects induced by knot placement in the spline representation, as well as the inherently noisy nature of the observations of each physical activity profile in the right tail, these parameters will be displayed and interpreted over the restricted interval $t \in [0, 0.98]$.  For the linear components $\hat{\alpha}_{\hat{\boldsymbol{\theta}}}(t)$ and $\hat{\boldsymbol{\beta}}_{\hat{\boldsymbol{\theta}}}(t)$, the interpretations are aided by the use global confidence bands using the survey boostrap methodology described in Section~\ref{ss:boot}.



\subsubsection{Linear Effect of Healthy Eating Index.}
\label{subsec:HEI_interpret}

\begin{figure}[t]
	\centering	\includegraphics[width=0.6\linewidth]{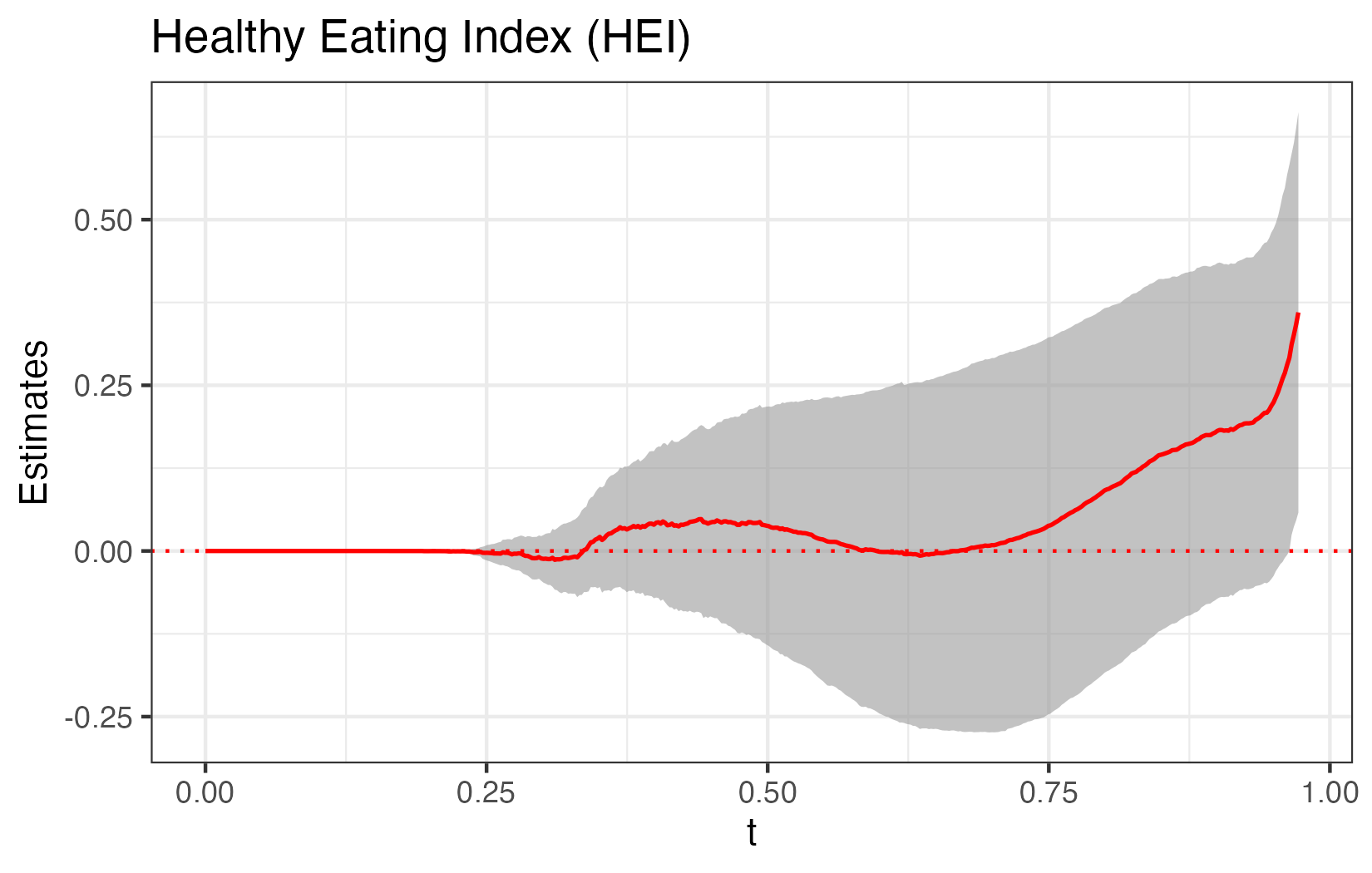}
	\caption{\footnotesize The estimated functional coefficient for the covariate HEI in the PL-FSI model \eqref{eq:wass_regression_basis} is shown here as the solid red curve and its $95\%$ global confidence band is indicated by the grey shaded region. The dotted red line at $0$ is for reference.}
	\label{fig:effects_compare_HEI}
\end{figure}

The Healthy Eating Index (HEI) reflects the diet quality of the participants. The outcomes of this analysis provide the statistical association of the diet patterns with physical activity patterns, with the corresponding coefficient function estimate being plotted in Figure \ref{fig:effects_compare_HEI}. For $t<0.25$, the estimated coefficient takes the constant value zero, reflecting the inactive portion of each participant's physical activity distribution. Even beyond the inactive region of quantile levels, HEI does not exhibit strong relevance for low to moderately high physical activity intensities ($0.25 < t<0.90$). However, for individuals with very high physical activity intensities (quantiles $t>0.90$), the plot suggests that 
healthier diet quality (higher HEI), is associated, on average, with 
greater physical activity intensities, after accounting for the other covariates in the model.
These findings underscore the intricate relationship between diet quality, lifestyles \cite{patterson1994health, leroux2015adult} and physical activity \cite{scarmeas2009physical}, especially in contexts where a prevalence of individuals actively engage in rigorous exercise regimens, such as high-intensity cardio and/or resistance training and/or physically-demanding manual labor. An interesting point is that establishing directionality, rather than merely interpreting statistical associations, would be ideal. However, to rigorously achieve this, we would require a mediation model \cite{feng2021role} within metric spaces using the 2-Wasserstein distance.


\subsubsection{Associations with Categorical Predictors: Ethnicity and Sex.}
\label{subsec:interpret_categorical}

The inclusion of categorical covariates as predictors of physical activity patterns is critical, as this allows for the identification of potential disparities in physical activity across various subgroups of the American population, opening the possibility for the development of targeted clinical interventions. The subsequent interpretations of the PL-FSI model fits are designed to address the following inquiries regarding significant epidemiological and public health questions related to differences across sexes and ethnicities.

\begin{figure}[t]
	\centering	\includegraphics[width=0.85\linewidth]{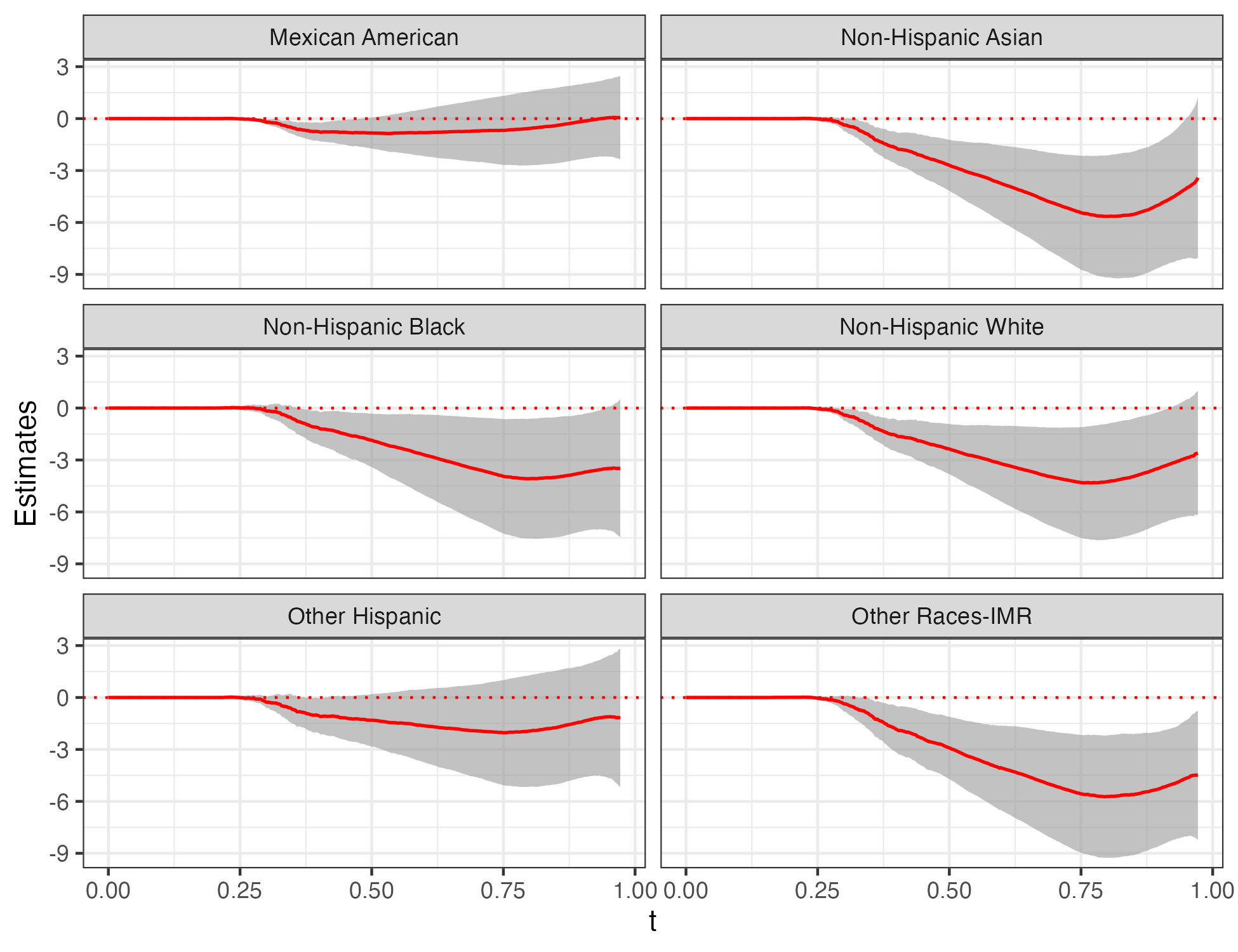}
	\caption{\footnotesize The model intercepts for the PL-FSI model \eqref{eq:wass_regression_basis}, are considered for male and female participants of different ethnic backgrounds. The intercepts for the males are subtracted from the intercepts of the females for each ethnicity, considering the numeric variables HEI, Age and BMI as fixed. The respective estimated parameter combinations are computed along with their 95\% global Confidence Intervals, and plotted as solid red curves and grey shaded regions respectively. The dotted red line at $0$ is for reference. The differences are considered for the ethnicities: Mexican American, Other Hispanic, Non-Hispanic White, Non-Hispanic Black, Non-Hispanic Asian, Other races including Multi-Racial.}
	\label{fig:effects_compare_MaleFemale}
\end{figure}

\begin{enumerate}
    \item Are there variations in physical activity levels between men and women, and how do these variations differ among the represented ethnicities?
    \item Do women of different ethnicities exhibit differences in physical activity levels, and if so, how do these differences vary?
    \item Do men of different ethnicities demonstrate disparities in physical activity levels, and if so, how do these differences vary?
\end{enumerate}

To address question 1, Figure \ref{fig:effects_compare_MaleFemale} illustrates the estimated model beta functions for male participants subtracted from those for female participants within each ethnic group. 
The estimated differences plotted as functions of $t$ for each ethnicity, along with their 95\% global confidence intervals, reveal that, for quantile levels $t$ around $0.30$ to $0.98$, men exhibit significantly higher physical activity levels, on average, than women among White, Black, Asian, and Other Races, including Multi-Racial ethnicities. However, in Mexican American and Other Hispanic groups, men and women display statistically similar levels of physical activity, on average.

\begin{figure}[t]
	\centering	\includegraphics[width=0.90\linewidth]{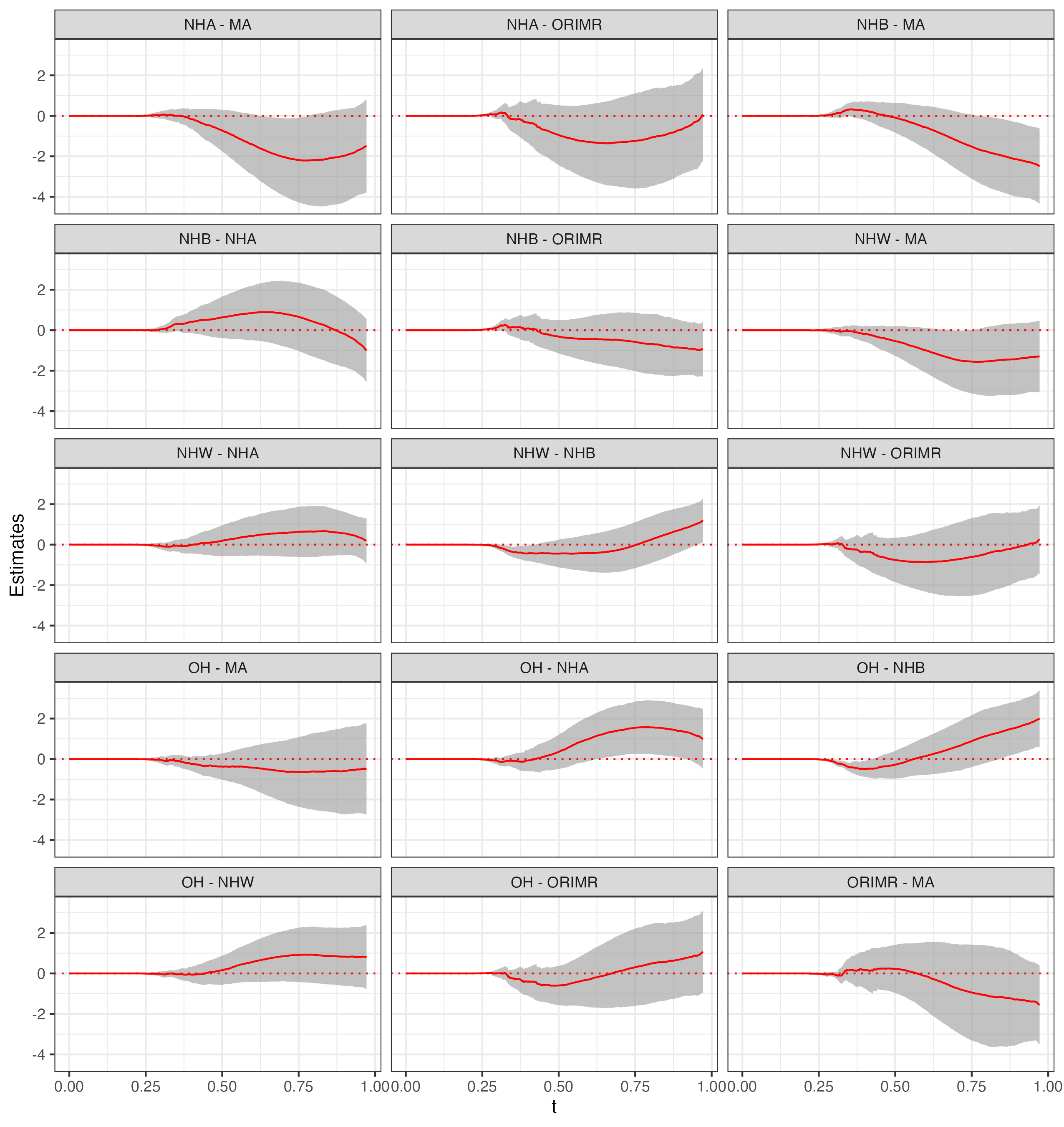}
	\caption{\footnotesize The estimated model intercepts for the PL-FSI model are compared for females of different ethnic backgrounds. The pairwise differences of such intercepts are displayed, along with their 95\% Confidence Intervals, as solid red curves and grey shaded regions respectively, considering the numeric variables HEI, Age and BMI as fixed. The dotted red line at $0$ is for reference. The title for each panel indicates the order of the differences of the intercepts. The abbreviations for the ethnicities are, OH: Other Hispanic, MA: Mexican American, NHW: Non-Hispanic White, NHB: Non-Hispanic Black, NHA: Non-Hispanic Asian, and ORIMR: Other Races Including Multi-Racial. As an example, the title 'OH - MA' indicates that the estimated intercepts for females identifying as Mexican American were subtracted from the estimated intercepts for females identifying as Other Hispanic ethnicity.}
	\label{fig:effects_compare_among_Females}
\end{figure}

\begin{figure}[t]
	\centering	\includegraphics[width=.90\linewidth]{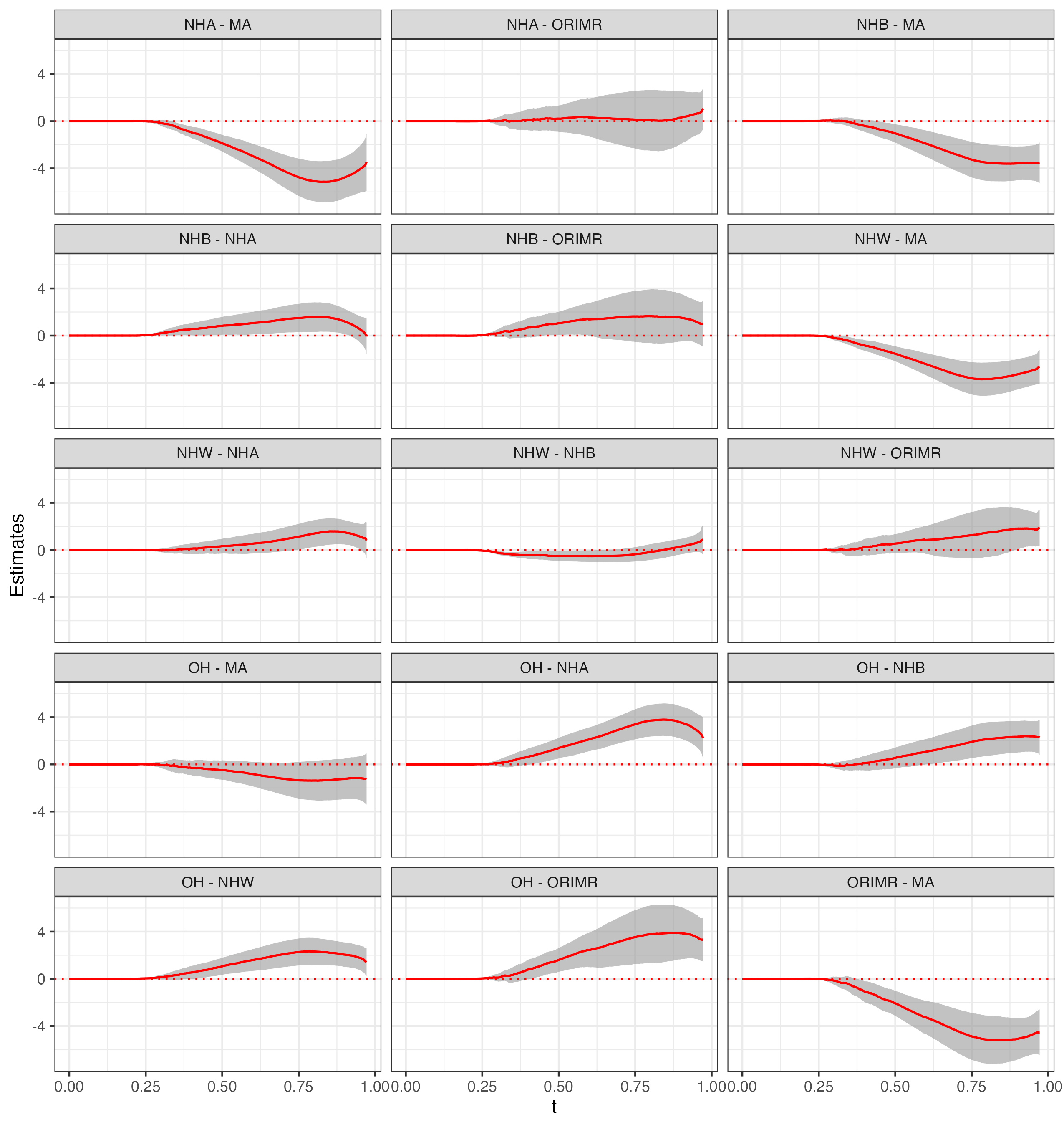}
	\caption{\footnotesize Estimated model intercepts for the PL-FSI model are compared for males of different ethnic backgrounds, conditional on fixed HEI, Age and BMI. The pairwise differences of such intercepts are computed along with their 95\% Confidence Intervals, and plotted here as solid red curves and grey shaded regions respectively.
 The dotted red line at $0$ is for reference. The abbreviations for ethnicities as well as the order of the differences are the same as in Figure \ref{fig:effects_compare_among_Females}.}
	\label{fig:effects_compare_among_Males}
\end{figure}

Addressing question 2 above, Figure \ref{fig:effects_compare_among_Females} illustrates pairwise differences in estimated model intercepts for females across different ethnicities, along with 95\% pointwise confidence intervals. These comparisons condition on fixed Age, HEI and BMI. Over the range of quantile levels $t$ from $0.30$ to $0.98$, Mexican American and Other Hispanic women, on average, showcase higher physical activity levels compared to certain other ethnicities. Women of White, Black, Asian, and Other Races, including Multi-Racial backgrounds, exhibit statistically similar physical activity levels, on average. Similarly, Mexican American and Other Hispanic women, on average, display comparable levels of physical activity.

Correspondingly, to address the question 3, Figure \ref{fig:effects_compare_among_Males} presents results akin to Figure \ref{fig:effects_compare_among_Females}, focusing on male participants. These comparisons condition on fixed Age, HEI and BMI. The estimates indicate that, for low to high physical activity levels ($t \in [0.30, 0.98]$), males identified as Mexican American or Other Hispanic ethnicities are more physically active, on average, compared to males of Black, White, Asian, and Other Races (including Multi-Racial backgrounds). While there is scientific evidence highlighting disparities in physical activity levels among different ages, sexes, and ethnicities \cite{cite-key, doi:10.1016/j.jacc.2023.12.019} in U.S. populations, to the best of the authors' knowledge, this is the first time that these disparities are addressed across the full range of accelerometer intensities using a distributional representation approach.
The model fit also suggests that men of White and Black ethnicities exhibit slightly higher physical activity levels than Asian men for certain quantile values. Men of Other Races, including Multi-Racial individuals, show slightly lower activity levels, on average, compared to Black and White men, but statistically similar levels to Asian men.

From a public health perspective, this underscores the need for tailored interventions to promote physical activity based on accelerometer data, considering variations among sexes and ethnicities. Such targeted strategies are pivotal in enhancing public health outcomes and addressing disparities. Public health policies aimed at promoting health should not be uniform across all groups, differing, for instance, between Asian men and Asian women or across ethnicities like Mexican American and Asian men, when conditioning on the same HEI, Age and BMI values. It is crucial to note that conducting sex-stratified analyses enhances the reliability of the presented findings as they reveal important interactions between these categorical variables when associated with physical activity intensities, when HEI, Age and BMI are the same.


\subsubsection{Nonlinear Associations with Age and BMI.}
\label{subsec:nonlinear_interpret}

Due to the semiparametric model structure, the intepretation of estimated associations between physical activity levels and the single index composed of Age and BMI requires some care.  As already demonstrated, this increased complexity yields a measurably improved model fit and predictive capacity.  In addition, by appropriately examining the estimated model components, an intuitive but nuanced association emerges. To begin, the estimated index parameter was $\hat{\boldsymbol{\theta}}=(0.2661, 0.9639)$ for the standardized variables BMI and Age respectively.  The nonparametric function \( g \) cannot distinguish between age and BMI since it only accounts for their linear combination as defined by the single index model. However, since both variables have been scaled to have a standard deviation of 1, the relative magnitudes of \(\hat{\boldsymbol{\theta}}\) do indeed reflect the relative contributions of each variable. In this case, age has a higher coefficient, suggesting a stronger association with the response variable compared to BMI.

As the effect is nonlinear, each of the four panels in Figure $2$ plots the 
fitted quantile values ${Y}^*(t,\boldsymbol{z}, \tilde{\boldsymbol{x}})$ for quantile levels $t =0.50, 0.75, 0.90,$ and $0.97$; here, the linear covariate $\boldsymbol{z}$ was fixed to represent the reference groups for sex and ethnicity and the median value of HEI, while $\tilde{\boldsymbol{x}}$ represents the standardized value of the Age and BMI combinations present along the horizontal and vertical axes of each panel.  The choice of displayed quantile levels, 
reflects the finding that the single index only exhibits a notable association with physical activity levels near or above the median intensities.  

For the lowest  BMI group ($<20$), people in the age range 55--70 are estimated by the model to perform the largest physical activity levels in terms of the median and third quartile ($t=0.50,\,0.75$).  However, for the same BMI range, people in the age range 25--35 perform the highest physical activity in the extreme right tail corresponding to quantiles $t=0.90, 0.97$. In each of the panels (or, quantiles) the age range for highest physical activity linearly decreases with increase in BMI. For the highest BMI in our study ($\approx 40$), the highest average physical activity for quantiles $t=0.50,\,0.75$ is estimated to occur within age range 40--55. However, in the quantiles $t=0.90,\,0.97$, the highest physical activity are shown by the BMI range 25--30 in the age range 20--25. Hence, these panels indicate that the nonlinear association, of average physical activity intensity, with Age and BMI is more pronounced for larger values of $t$. This estimated association is more robust for middle age individuals with intermediate BMI values than in the rest of the range. 
As the age increases and the BMI decreases, individuals are more likely to have lower activity intensities, especially in the extremes of Age and/or BMI.

\begin{figure}[t]
	\centering
	\includegraphics[width=.496\linewidth]{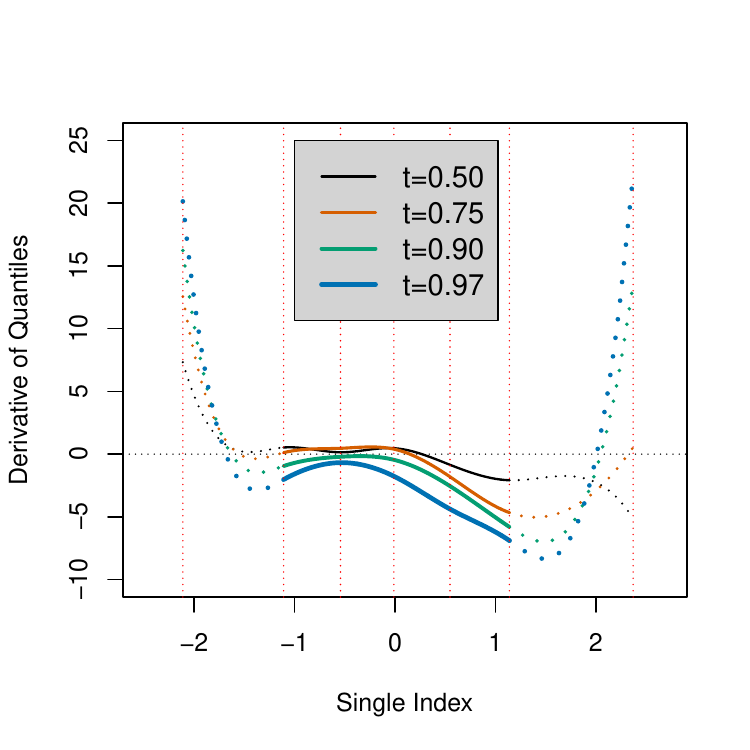}
\caption{\footnotesize (Left) Plot of $(\partial \hat{g}/\partial u) (u, t)$, the derivative of the spline-based estimate of the nonlinear component defined in \eqref{eq:NLFitDer}, for $u$ across the empirical range of observed index values $\hat{\boldsymbol{\theta}}^{T}\boldsymbol{X}_i,$ $i = 1,\ldots,n,$ and for quantile levels $t=0.50$ (black), $0.75$ (red), $0.90$(green),and $0.97$ (blue). Knot locations are shown as vertical dotted red lines, and the derivative curves are depicted as solid (respectively, dotted) within (resp., outside of) the interior knot range. (Right) Histogram of observed index values $\hat{\boldsymbol{\theta}}^{T}\boldsymbol{X}_i,$ $i = 1,\ldots,n.$	(Right) Histogram of the distribution of the single index projection evaluation for 4,616 included participants in NHANES 2011-2014
\label{fig:graphderv2}}
\end{figure}

As an additional visualization of the nonlinear association with these covariates in the model, consider a direct analysis of the derivative of the nonlinear fit, namely
\begin{equation}
    \label{eq:NLFitDer}
    \frac{\partial \hat{g}}{\partial u}(u, t) = \sum_{k = 1}^{K+s}\hat{\gamma}_{\hat{\boldsymbol{\theta}},k}(t)\left[\frac{d}{{d} u} \phi_k(u)\right].
\end{equation}
The reason to consider the derivative is that, if $\hat{g}$ were a linear function in $u$, then this derivative could be interpreted in the same way as the linear coefficient estimates in the previous sections.  In other words, it is the direct counterpart of the linear coefficient functions for this nonlinear term. The left panel of Figure \ref{fig:graphderv2} displays the behavior of \eqref{eq:Y_est_notL2} across the relevant range of values $u$, with respect to the empirical values $\hat{\boldsymbol{\theta}}^{T}\boldsymbol{X}_i$ (whose distribution is depicted via a histogram in the right panel of the figure) that were used to generate the estimates, for $t = 0.5, 0.75, 0.9, 0.97$. As the observed shape of the curves within the interior knots is most reliable, this region is indicated by the solid portion of each curve in the figure. The various derivative curves suggest that, given the covariates in the linear term, for negative values of the single index containing BMI and Age, there is little to no association with the physical activity quantile response; for positive single index values, the estimated association becomes negative. Since both elements of $\hat{\boldsymbol{\theta}}$ are positive, these findings imply that the model reflects a negative association between physical activity quantiles and BMI/Age when at least one of these is large. Furthermore, the strength of this negative association increases as the quantile level $t$ becomes larger, as evidenced by the increasingly negative derivative estimates in the left panel as $t$ increases. For instance, the derivative for $t = 0.5$ (the median physical activity quantile) is only slightly negative for values of $u$ near zero, whereas it steadily decreases as one examines the curves for $t = 0.75, 0.9, 0.97$. In short, for positive values of $u = \hat{\boldsymbol{\theta}}^T\boldsymbol{x}$, average physical activity quantiles above the median tend to decrease, and the rate of decrease becomes more pronounced for both larger values of $u$ and quantile level $t$.



\section{Discussion}
 \label{sec:disc}

The main contribution of this paper is to propose a new PL-FSI regression model to analyze responses of a distributional functional nature. The new methods have been implemented to analyze the physical activity data from the NHANES database 2011-2014, for participants aged $20 -80$ and in BMI range $18.5 -40$. The NHANES survey weights were incorporated into the PL-FSI algorithm using the sampling mechanism of the Horvitz-Thompson type estimator \cite{kish1965survey} to construct a weighted least squares criterion to estimate the model parameters.

The findings from application of this new model are summarized as follows.
\begin{enumerate}
\item  The discrepancies in physical activity levels between men and women of different ethnicities were examined in the American population. Associations between physical activity and the continuous variables HEI, BMI, and Age were quantified, visualized, and interpreted across the range of human physical activity intensities, as opposed to just the mean or median activity level, due to the new quantile distributional representation of physical activity. For example, it was shown that diet is important only in the high-intensity levels of physical activity range; a better diet, according to the HEI score, is related to more exercise. We also show that the Mexican American and Other Hispanic groups are the most active individuals in the American population for both men and women. An interaction between Age and BMI was discovered and exploited in determining their association with energetic expenditure, valid more specifically in the moderate to higher intensities of physical activity levels.
\item  The modeling advantages of the new PL-FSI algorithm over the classical global Fréchet regression model were shown in terms of adjusted Fréchet R-squared and mean square prediction error.  In addition, interpretation of the nonlinear term in the PL-FSI model was demonstrated using the gradient of a conditional mean function.
\end{enumerate}


From a practical perspective, these new results illustrate the variation in physical activity across the range of accelerometer intensities, unlike previous models that focus on scalar summaries and averages \cite{leroux2019organizing}.
The results derived from the PL-FSI model, primarily based on demographic variables, show that we can define expected physical activity levels in different U.S. populations. From a public health perspective, this approach can be generalized—for instance, by proposing tolerance regions for physical activity and creating new recommendations about expected physical activity levels, following \cite{matabuena2024conformal}. However, the model in this previous work does not provide interpretable statistical associations as our case, as they are no longer explainable.

From a methodological point of view, we propose the first PL-FSI regression model in the context of object data analysis to bridge the gap between the global Fréchet regression and the Frechet single index model, while preserving the interpretability of the predictors and parameter estimates. To the best of our knowledge, this is also the first regression model to incorporate survey data in the context of object data analysis.  

The most popular approach to analyzing accelerometer data is through finite dimensional compositional metrics. Here, the functional extension of these metrics \cite{10.1093/jrsssc/qlad007}, was instead used to capture more information about physical activity from an individual by adopting the mathematical framework of the $L^2$-Wasserstein space. 
Due to the positive probability at zero physical activity level for each individual (corresponding to periods of inactivity), the quantile function, which is intimately connected with the Wasserstein metric, provides a natural functional representation of such mixed distributions. In addition, the range of values measured by the accelerometer varies widely among individuals and groups, which can present difficulties when trying to apply the standard distributional data analysis methods in this setting \cite{10.1093/jrsssc/qlad007}. For example, functional compositional transformations can be an alternative strategy to creating a regression model about physical activity in a linear
space \cite{van2014bayes, petersen2016functional, hron2016simplicial}. However, the distributional physical activity representation arises from a mixed-stochastic process (see Figures \ref{fig:raw}, \ref{fig:raw2} for more details) that prevents the naive use of the linear functional data methods that typically utilize a basis of smooth functions to represent the functional response and/or the functional parameters in the model, due to the discontinuity of the quantile function in the transition of the inactivity to activity in the physical exercise. While specialized basis functions that allow for jumps in the functional parameters or their derivatives could be used, the proposed model and its estimation procedure demonstrate that it is not necessary to do so.
As future work, we propose generalizing the distributional variable selection model proposed in \cite{coulter2024fast} for survey data, and extending our PL-FSI semi-parametric approach to select the most relevant predictors in settings with a larger pool of variables. Additionally, providing prediction regions could be highly valuable for modeling scientific problems. Leveraging a previous framework for uncertainty quantification in metric spaces \cite{matabuena2024conformal}, adapted for survey data and applied to the PL-FSI model, can be especially relevant. For instance, defining tolerance regions from a distributional perspective on physical activity could address emerging scientific challenges that are currently focused on scalar variables as step counts in the literature.

Missing data  is another  significant challenge in wearable data analysis, especially when studying younger populations or during shorter monitoring periods. Although accelerometers generally provide more consistent data quality, smartphone-based tracking of physical activity often introduces greater variability and data gaps, which complicate the reliability of analyses. Addressing these issues necessitates the development of innovative methods and stringent criteria specifically tailored to smartphone-derived data to ensure data integrity. Implementing these approaches is essential for producing accurate and robust results when applying distributional representations of physical activity as both predictor and response variables.

`
The analysis of complex statistical objects in biomedical science provides an excellent opportunity to create new clinical biomarkers that enrich those available for medical decision-making beyond those commonly used to monitor the health and evolution of diseases. For example, distributional representations are a significant advance in digital medicine as a digital biomarker. However, the generality of techniques introduced also enables the application of the methods developed here to other complex statistical objects such as connectivity graphs, shapes, and directional objects. These methods have potential to introduce new clinical findings in a broad list of clinical situations, for example in neuroimaging and in phylogenetic tree analysis \cite{relion2019network,zhou2021dynamic,nye2017principal,dubey2022modeling,yuan:12}.  Furthermore, with the increasing availability of data from large cohort studies,
such as from longitudinal surveys with carefully designed subpopulation sampling weights, the methods provided here will gain more popularity among practitioners. The use of complex statistical objects will undoubtedly enhance daily statistical practice in biomedical applications.

\end{document}